\documentstyle[aps,prb,amsfonts,amssymb,twocolumn,epsf]{revtex}

\begin{document}

\draft

\title{Is the Multichannel Kondo Model  Appropriate to Describe the 
Single Electron Transistor?}
\author{Gergely Zar\'and,$^{1,2}$  Gergely T. Zim\'anyi,$^2$ and Frank Wilhelm$^3$}
\address{${}^1$Institute of Physics, Technical University of
Budapest, H 1521 Budafoki \'ut 8., Budapest, Hungary \\
${}^2$UC Davis, 1 Shields Ave. CA 95616 \\
${}^3$Institut f\"ur Theoretische Festk\"orperphysik, 
Universit\"at   Karlsruhe, 76128 Karlsruhe, Germany}

\twocolumn[\hsize\textwidth\columnwidth\hsize\csname
@twocolumnfalse\endcsname%
\date{today}
\maketitle 

\begin{abstract}
We investigate the low-temperature dynamics of single 
electron boxes and transistors
 close to their  degeneracy point using renormalization 
group methods.  We show that intermode scattering is a relevant  
perturbation and always drives the system to the two-channel Kondo 
fixed point, where the two channels correspond to the real spins of the
conduction electrons.  However, the crossover temperature $T^*$, 
below which Matveev's two-channel Kondo scenario 
${\rm [}$K.A. Matveev, Phys. Rev. B {\bf 51}, 1743 (1995)${\rm ]}$ develops decreases 
exponentially with the number of conduction modes in the tunneling 
junctions and is extremely  small in most cases. Above $T^*$ the 
'infinite channel model' of  Ref.~\protect{\onlinecite{Schoen}} 
turns out to be a rather good approximation.
We  discuss the experimental limitations and  suggest a new experimental 
setup to observe the multichannel Kondo behavior. 
\end{abstract}
\vskip0.0pc
]

\section{Introduction}

\begin{figure}
\epsfxsize=5cm
\begin{center}
\epsfbox{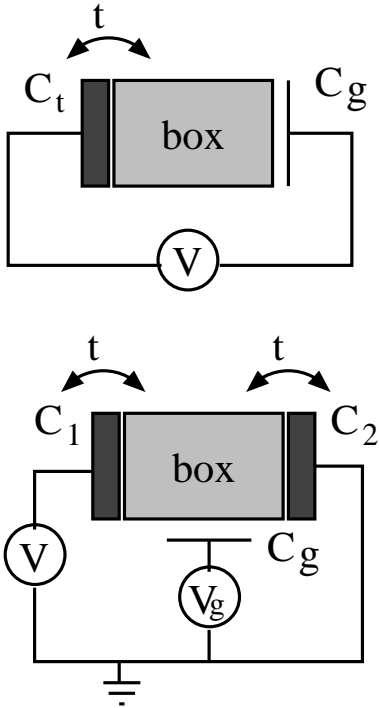}
\end{center}
\caption{\label{fig:set}
Sketch of the single electron box (SEB) and the 
single electron transistor (SET).}
\end{figure}

The single electron box (SEB) and the single electron transistor (SET) 
are  basic elements of mesoscopic devices and have been  
studied extensively.\cite{setref,DevoretLesHouches} Both consist 
of a single small metallic or semiconducting box connected to 
one (single electron box, or   SEB) or two (single electron 
transistor, or SET) leads. Additionally, in the SET
 a gate electrode is attached to 
the box to control the actual charge on the dot (see Fig.~\ref{fig:set}).

The electrostatic energy of the box is well-described by the 
classical expression\cite{setref} 
\begin{equation}
H_C = {e^2\over 2C} \left( n_{\rm box}  - {V_g C_g\over e}\right)^2\;,
\end{equation}
where $C$ denotes the capacitance of the island, $C_g$ is the gate 
capacitance, $e$ is the electric charge, $V_g$ stands for the gate 
voltage, and $ n_{\rm box}$ is the number of extra electrons on the 
island.  For   box sizes in the $0.1\mu {\rm m}$ range the capacitance 
$C$ of the box can be  small enough so that the charging energy 
$E_C = e^2/2C$ associated with putting an extra electron 
on the island can safely be around  ${\rm mV}$ range. Therefore, 
unless $N_g = {V_g C_g\over e}$ is a half-integer, it costs
a finite energy to charge the island, and at low enough temperatures 
 the number of electrons  on the island becomes 
quantized and 
a Coulomb blockade develops provided the quantum fluctuations are
not too strong. In this Coulomb blockade regime the transport through 
the island is suppressed.

The situation is dramatically different for dimensionless gate voltages 
$N_g = {V_g C_g\over e} \approx n+1/2$. For $N_g = n+1/2$ the two states
$n_{\rm box} = n$ and $n_{\rm box} = n+1$ become degenerate, and quantum 
fluctuations between the island and the leads become important.
Assuming that the mean level spacing on the island is smaller than any 
energy scale (temperature, $E_C$, etc.) two scenarios 
have been  suggested: (a) It has been proved  
by Matveev\cite{MatveevJETP,MatveevPRB}  that if 
the leads are connected to the island via  a  {\em single} conduction mode 
then --- close to the degeneracy point and at low enough temperature ---
the physics of the SET (SEB) becomes  identical to that of the 
two-channel Kondo model. Indeed, the fingerprints of the two-channel 
Kondo  behavior 
have been  observed recently on  semiconducting single electron 
transistors.\cite{Berman}  (b) In the opposite limit one assumes that 
the tunneling to the island happens through an infinite number of 
identical modes.\cite{Schoen,KoenigSchoellerSchoen} 
This model has been applied very successfully for the description 
of metallic islands.\cite{Devoret}
The predictions of this infinite channel model 
are, however,  very different from those of the two-channel Kondo model: 
the conductance of the SET, for instance, scales to zero as 
$\sim T$ in the two-channel Kondo picture,\cite{MatveevPRB} while it 
is proportional to $1/\ln(E_C/T)$ in the infinite channel 
scenario.\cite{Schoen}

The purpose of the present paper is to treat the general case of
{\em finite} conduction modes in the lead and to reconcile the 
apparent contradiction between the two pictures above. 
We show that both models
 capture the physical  properties of the SEB (SET), however, 
they are appropriate in very {\em different regimes}. Carrying out a
renormalization group  analysis  we show  that there exists a 
crossover energy $T^*$. Above $T^*$, even for 
$N \sim  20$ tunneling modes the system is well characterized  
by the conductance of the tunnel junctions and the SEB (SET)  
is satisfactorily  described by the $2N$-channel model of 
Ref.~\onlinecite{Schoen}. Nevertheless,  for small mode numbers 
pronounced deviations occur, and 
similar deviations  appear for larger values of $N$ in the presence 
of {\em pinholes} in the junction, which offer a plausible explanation 
to the  deviations observed in Ref.~\onlinecite{Devoret}.

Below $T^*$, on the other hand, the detailed structure of the 
tunneling matrix becomes important: At very low $T$  only a single 
conductance mode dominates the physics and a two-channel
Kondo effect develops. Unfortunately, in most situations 
 $T^*$ (and thus the Kondo temperature $T_K\ll T^*$) turns out to be
extremely small, and the two-channel Kondo physics cannot be
observed. In fact, very special  experimental setups are needed to 
observe  the two-channel Kondo behavior, as we shall discuss it in detail 
in our concluding section.

The paper is organized as follows: 
In Sec.~\ref{sec:model} we describe the models applied.
Secs.~\ref{sec:seb}  and \ref{sec:set} are devoted to the  
analysis of the single  electron box and the single electron 
transistor, respectively. 
Finally, in Sec.~\ref{Conclusions} we discuss the possibility 
of experimental observations of the low-energy Kondo-like 
behavior  and summarize our conclusions.

\section{The Models}
\label{sec:model}

\subsection{Hamiltonian of the SEB}

For the sake of simplicity, let us first concentrate on the single 
electron box  and   generalize our results to the SET later on. 
Usually, the lead is described by means of $N$ independent 
non-interacting one-dimensional electron modes: 
\begin{equation}
H_{\rm lead} = \sum_{n = 1}^N \sum_{\epsilon,\sigma}  \epsilon \; 
c^\dagger_{\epsilon n \sigma; {\rm lead} }
c_{\epsilon n \sigma; {\rm lead} }\;, 
\end{equation}
where $c^\dagger_{\epsilon n \sigma; {\rm lead} }$ crates an electron
on the box with spin $\sigma$, mode index $m$ and energy $\epsilon$.
(Note that to avoid confusion, we do not follow  the usual terminology and 
use deliberately the expression  {\em conduction  mode} instead of the 
wording 'conductance channels', more frequent  in the literature.)

In the present work we assume that the level spacing $\Delta$ at the 
island is much smaller than any energy scale in the problem, and 
therefore these discrete  levels may be represented as
 a single particle continuum on the island.
This assumption is crucial to obtain the Kondo physics discussed
in this paper, since the level spacing provides an infrared cut-off 
which ultimately kills the logarithmic singularities and the Kondo 
effect. Based on these assumptions  we express the Hamiltonian of the 
island as\cite{setref}
\begin{eqnarray}
H_{\rm box} & = & \sum_{m = 1}^M \sum_\epsilon \epsilon\; 
c^\dagger_{\epsilon m \sigma; {\rm box} } c_{\epsilon m \sigma; 
{\rm box} } \nonumber \\
&+&  E_C \left( n_{\rm box}  - {V_g C_g/ e}
\right)^2  \;.
\label{eq:box}
\end{eqnarray}
In Eq.~(\ref{eq:box})  we assumed $M$ independent modes 
on the box,\cite{footnote} and defined the number 
$n_{\rm box}$ of electrons on the island as
\begin{equation}
n_{\rm box}  = \sum_{m = 1}^M \sum_{\epsilon,\sigma}  
:c^\dagger_{\epsilon m \sigma; {\rm box} } c_{\epsilon m \sigma; 
{\rm box}}:\;\;, 
\label{eq:n_box}
\end{equation}
where the symbol $:...:$ denotes normal ordering.

As usually, in Eq.~(\ref{eq:box})  we implicitly used the  
assumption  that the collective charge excitations decouple from the single 
particle excitations\cite{DevoretLesHouches} and that the 
electron-electron interaction can be fully taken  into account by 
the classical Coulomb  interaction term.  The validity of this 
approximation  relies heavily on the fact the collective charge 
excitations relax  extremely fast  compared to all other time scales 
involved.

The coupling of the box  to the lead is described by 
a standard tunneling Hamiltonian:
\begin{equation}
H_{\rm tun} = \sum_{n=1}^N \sum_{m=1}^M 
\sum_{\epsilon,\epsilon',\sigma} 
\left( T_{mn} c^\dagger_{\epsilon m \sigma; {\rm box} } 
c_{\epsilon' n \sigma; {\rm lead} } + 
\mbox{h.c.}\right)\;,
\label{eq:H_tun}
\end{equation}
where we neglected the energy dependence of the elements of the 
$M\times N$ tunneling matrix  $T_{mn}$

It is very important that the tunneling is  
{\em diagonal} in the spin indices, however, it is generally 
{\em non-diagonal} in the mode indices. As we shall see later, 
the twofold spin degeneracy  is the basic origin 
of the very low-temperature {\em two-channel} Kondo 
effect.\cite{MatveevJETP} 
Once magnetic field is applied the symmetry between spin up and 
spin down conduction electrons  is broken, and the 
system flows to the single channel Kondo fixed point.

The tunneling Hamiltonian of the SET differs only slightly 
from that of the SEB. In this case there are two leads that are 
connected to the island. However, as first shown by Averin and 
Nazarov,\cite{AverinNazarov} at temperatures larger than the level spacing 
coherent processes connecting the two leads are strongly 
suppressed.  Therefore, one can formally separate from each-other
those single particle
states on the island which participate in the tunneling from the 
first and the second lead, respectively.\cite{MatveevPRB}  
These tunneling processes are then only 
correlated by the  very fast Coulomb interaction which allows for the presence 
of only one   excess electron  on the island.

Thus for the SET the effective  Hamiltonian of the island becomes\cite{MatveevPRB}:
\begin{eqnarray}
H_{\rm box} & = & \sum_{m = 1}^M \sum_\epsilon \epsilon\;\bigl( 
{c^{(1)}}^\dagger_{\epsilon m \sigma; {\rm box} } 
{c^{(1)}}_{\epsilon m \sigma; {\rm box} } 
 \nonumber\\ 
& + & {c^{(2)}}^\dagger_{\epsilon m \sigma; {\rm box} } 
	{c^{(2)}}_{\epsilon m \sigma;  {\rm box} }
\bigr)
\nonumber \\
&+&  E_C \left( n^{(1)}_{\rm box} +  n^{(2)}_{\rm box}  - {V_g C_g/ e}
\right)^2  \;,
\nonumber
\end{eqnarray}
where the indices $(1)$ and $(2)$ refer to single particle states 
participating in the  tunneling from the first and second leads, 
respectively, and the number operators $n^{(1)}_{\rm box}$ and
$n^{(2)}_{\rm box}$ are defined similarly to Eq.~(\ref{eq:n_box}).

The tunneling Hamiltonian of the SET reads:
\begin{equation}
H_{\rm tun} = \sum_{\scriptstyle f=1,2 \atop 
\scriptstyle n, m, \epsilon,\epsilon',\sigma} 
\left( T^{(f)}_{mn} {c^{(f)}}^\dagger_{\epsilon m \sigma; {\rm box} } 
{c^{(f)}}_{\epsilon' n \sigma; {\rm lead} } + 
\mbox{h.c.}\right)\;,
\end{equation}
where the index $f$ refers to the two junctions. For the sake of simplicity, 
we assumed that the number of modes in the two leads ($N^{(1)}$ and  $N^{(2)}$) 
and the  number of tunneling modes on the island ($M^{(1)}$ and $M^{(2)}$ ) 
is identical for both  junctions($N^{(1)} = N^{(2)} = N$  
and $M^{(1)} = M^{(2)} = M$ ). This simplification does not 
modify our  results because the two tunneling matrices $T^{(1)}_{mn}$ 
and  $ T^{(2)}_{mn}$ are assumed to be  completely uncorrelated.

In the following we focus to  the vicinity 
of the degeneracy points, $V C_G/e = n + 1/2$. As already mentioned in the 
introduction at these gate voltages  the charge states $n_{\rm box} = n$ 
and $n_{\rm box} = n +1$  become  degenerate and quantum fluctuations 
dominate. For temperatures (energy scales) below the charging energy 
$E_C = e^2/2C$ one can safely project out all the other charging states, 
represent the 
two states $n_{\rm box}  = n +1/2 \pm 1/2$ as  two states of a pseudospin
$S_z  = \pm 1/2$, and rewrite the tunneling part of 
SEB Hamiltonian in the following form:
\begin{eqnarray}
H &=& \sum_{\epsilon,\sigma,\alpha }  \sum_{n = 1}^{N_\alpha}  
\epsilon \; c^\dagger_{\epsilon n \sigma; \alpha }
c_{\epsilon n \sigma; \alpha } - hS_z 
\label{eq:H_restr} \\
& + & \sum_{\scriptstyle \epsilon,\epsilon',\sigma \atop 
\alpha,\beta,n_\alpha, n_\beta} 
 ( c^\dagger_{\epsilon n_\alpha \sigma; \alpha } 
T_{n_\alpha,m_\beta} \sigma^+_{\alpha\beta} S^- 
c_{\epsilon' m_\beta \sigma; {\beta}} +{\rm  h.c.} 
)\;,  
\nonumber 
\end{eqnarray}
where the 'orbital pseudospins' $\alpha,\beta = ({\rm 'box'},\;{\rm
'lead'})$ indicate the position of an electron and couple  to $S$, 
$\sigma^\pm = \sigma_x \pm i \sigma_y$ denote  Pauli 
matrices, and  the index $n_\alpha$ takes values  $n_\alpha = 1,..,N_{\alpha}$
($N_{\rm lead} = N$ and $N_{\rm box} = M$).
The effective field $h = e\delta V C_G/C$ measures the 
distance from the degeneracy point, with  $\delta V = V - e(n+1/2)/C_G$.

For the SET Eq.~(\ref{eq:H_restr}) gets modified in that 
the two leads provide two conduction electron 'channels'
($f=1,2$) coupled to the charge pseudospin: 
\begin{equation}
H_{\rm tun} =  \sum_{\scriptstyle f, \epsilon,\epsilon',\sigma \atop 
\alpha,\beta,n, m} 
 ( {c^{(f)}}^\dagger_{\epsilon n \sigma; \alpha } 
T^{(f)}_{n,m} \sigma^+_{\alpha\beta} S^- 
c^{(f)}_{\epsilon' m \sigma; {\beta}} +{\rm  h.c.} 
).  
\label{eq:H_restr_SET}
\end{equation}
Note that the 'channel' label $f$  has a role  essentially  different 
from that of the real spin of the  electrons: While there is a full SU(2) symmetry 
associated to the latter, the former is merely a conserved quantity 
(corresponding to a $U(1)$ symmetry only).

In the formulation above the case  $M=N=1$ corresponds 
to Matveev's two-channel Kondo model,\cite{MatveevJETP} while 
 in the limit  $M=N\to \infty$  and $T_{nm} = \delta_{nm} T\sim 
1/\sqrt{N}$ we recover the infinite channel model\cite{Schoen} 
mentioned in the 
Introduction. Obviously, both limits are somewhat specific: 
In many realistic systems $M,N>1$ and the first approximation seems 
to be inadequate.  The second approximation, on the other hand, 
contains an artificial  $SU(2N)$ symmetry: There is no reason for the 
tunneling matrix element  to be diagonal in the mode indices at all, 
and even more to have  identical matrix element in each tunneling mode. 
In fact,  any defect, roughness, etc. present in a real junction will 
produce cross-channel tunneling, and even the simplest models of 
a perfect tunnel junction with $N=M$ give different tunneling eigenvalues for 
the different tunneling modes. The philosophy behind this second 
approach is that the only physically relevant parameter is the conductance 
of the junction and therefore the artificial symmetry introduced has no 
effect.  As we shall see,  this philosophy is only partially justified: cross-mode 
tunneling --- breaking  this artificial $SU(2N)$ symmetry --- is in reality 
a {\em relevant } perturbation and leads the system ultimately 
to the two-channel Kondo fixed point.

\section{Perturbative scaling analysis of the SEB}
\label{sec:seb}
It has been shown longtime ago that the Hamiltonian of
Eq.~(\ref{eq:H_restr}) generates logarithmic singularities when 
perturbation theory in the tunneling amplitude
is developed.\cite{MatveevJETP} To deal  with these logarithmic singularities 
one has to  sum up the perturbation series 
up to infinite order. The easiest way to do this 
is by constructing the renormalization group (RG)  equations.
Fortunately, this straightforward but rather tedious calculation 
can be avoided by rewriting the Hamiltonian (\ref{eq:H_restr})
in the following form: 
\begin{equation}
H_{\rm int}=\sum_{i=1}^3 \sum_{\sigma,\epsilon,\epsilon^\prime,r,r^\prime}
 V^i_{r r^\prime}
\tau^i c^+_{\epsilon r \sigma} c_{\epsilon^\prime r^\prime
\sigma}\;,
\label{eq:TLS}
\end{equation}
and observing that Eq.~(\ref{eq:TLS}) is formally identical to the 
Hamiltonian of a non-commutative two-level system.\cite{TLSreviews}
The indices  $r$ and $r'$  in Eq.~(\ref{eq:TLS}) take the values 
$r,r' = (1,..,M+N)$, the $\tau^i$'s denote Pauli matrices ($\tau^i = 2S^i$),
 and the $V^i_{rr^\prime}$'s can be written in a block matrix 
notation as
\begin{eqnarray}
&& {\bf V}^x =
{1\over2}\left(\begin{array}{cc}
0  & {\bf T} \\
{\bf T}^\dagger & 0
\end{array}\right) \;,\\
&&{\bf V}^y =
{1\over2i}\left(\begin{array}{cc}
0  & - {\bf T} \\
{\bf T}^+ & 0
\end{array}\right) \;, \\
&&{\bf V}^z =
\left(\begin{array}{cc}
{\bf Q}  & 0 \\
0  & {\bf \tilde Q}
\end{array}\right) \;,
\label{eq:deviation}
\end{eqnarray}
where we introduced the tensor notation $T_{mn} \to {\bf T}$.
 The $M\times M$ and $N\times N$
Hermitian matrices ${\bf Q}$ and ${\bf \tilde Q}$ vanish in the 
bare Hamiltonian, but they  are dynamically generated under scaling. 
They correspond to charging state dependent back scattering off the 
tunnel junction:
\begin{eqnarray}
H_{\rm back}&=&\sum_{\epsilon,\epsilon',\sigma,\sigma'}
2S_z\;\bigl[\sum_{n,n' = 1}^N {\tilde Q}_{nn'} 
c^\dagger_{\epsilon n \sigma; {\rm lead} } c_{\epsilon' n' \sigma'; {\rm lead} } \\
&+&\sum_{m,m' = 1}^M  Q_{mm'} 
c^\dagger_{\epsilon m \sigma; {\rm box} } c_{\epsilon' m' \sigma'; {\rm box} }
\bigr]\;.
\end{eqnarray}
The scaling equations of the two-level system have been first derived
in Ref.~\onlinecite{VladZaw} and its possible fixed points and their stability 
have been carefully analyzed in Ref.~\onlinecite{ZarVlad}. 
Using this mapping we can easily construct the RG equations for the 
SEB: 
\begin{eqnarray}
{d{\bf t}\over dx}& = & 2({\bf t q} - {\bf \tilde q t})
\nonumber \\
& - &
2 {\bf t} \left[ {\rm Tr}\{ {\bf t t^\dagger}\} 
+ 2 {\rm Tr}\{ {\bf q  q }\}
+ 2 {\rm Tr}\{ {\bf \tilde q \tilde q }\}\right]\;,
\label{eq:tscaling}\\
{d{\bf \tilde q}\over dx} & = &{\bf t^\dagger t}  - 4 {\bf \tilde q}
{\rm Tr}\{ {\bf t t^\dagger}\} \;, 
\label{eq:qscaling}\\
{d{\bf  q}\over dx} &=& -{\bf  t t^\dagger}  - 4 {\bf  q}
{\rm Tr}\{ {\bf t t^\dagger}\}
\label{eq:qtildescaling}\;. 
\end{eqnarray}
Here we introduced the scaling variable $x = 
\ln(E_c/\max\{T,\omega,..\})$  and defined dimensionless couplings as
$t_{mn} \equiv (\varrho^{\rm box}_m \varrho^{\rm lead}_n)^{1/2}
T_{mn}$ (and similarly, $q_{mm'} 
\equiv (\varrho^{\rm box}_m \varrho^{\rm box}_{m'})^{1/2}
Q_{mm'}$, and $\tilde q_{nn'} \equiv (\varrho^{\rm lead}_n 
\varrho^{\rm lead}_{n'})^{1/2} \tilde Q_{nn'}$) with 
$\varrho^{\rm box}_m$  ($\varrho^{\rm lead}_n$) the density of states
at the Fermi energy in mode $m$ (mode $n$) of the box (lead). 

The scaling equation for the effective field $h$ can be obtained from
that of  the splitting in the two-level system problem:
\begin{equation}
{dh \over dx} = - 4 h {\rm Tr}\{ {\bf t t^\dagger}\}\;. 
\label{eq:hscaling}
\end{equation}

These scaling equations are appropriate provided $g\equiv 
{\rm Tr}\{ {\bf t
t^\dagger}\} < 1$, otherwise the perturbative RG breaks down. They
must be   solved with the initial condition ${\bf q}(x=0) = {\bf
\tilde q}(x=0) = 0$ and ${\bf t}(x=0) = {\bf t}^{\rm bare}$. Away
from the degeneracy point the ``magnetic field'', i.e. the deviation  from 
the degeneracy point is a relevant perturbation,  and the  scaling must be  cut off 
at an energy scale $h^*$ determined selfconsistently from the condition
$h(x=\ln(E_C/h^*)) = h^*$. 

Up to logarithmic accuracy, the conductivity can then be 
expressed in terms of  the scaled  
dimensionless  tunneling rate ${\bf t}(x=\ln[E_C/\max\{T,h^*\}])$ as
\begin{equation}
G(T) = G_0 {\rm Tr}\{ {\bf t}(x) {\bf t}^\dagger (x)\}
\end{equation}
where  $G_0 = 8\pi^2e^2/h$ denotes a  universal conductance 
unit.

The analogues of Eqs.~(\ref{eq:tscaling}-\ref{eq:qtildescaling})
have been analyzed very carefully in Refs.~\onlinecite{ZarVlad}, 
where it was shown  by means of a
systematic $1/N_s$ expansion ($N_s$ being the spin of the electrons) 
that they have a  unique stable fixed point, 
{\em identical to the two-channel 
Kondo fixed point}. At this latter two orbital quantum 
numbers prevail, and the others become irrelevant. In the present 
context this statement means that there will be a {\em unique} 
'effective tunneling mode' in the lead (it is some combination of the 
original tunneling  modes in the lead), and 
another {\em unique} 'tunneling mode' 
in the box (also a linear combination of the original 
modes in the box): The $T=0$ effective Hamiltonian of the model 
at the degeneracy point corresponds to tunneling between these two 
modes only,  and all the other modes can be neglected. This
theorem therefore justifies the use of Matveev's effective model 
at very low temperatures even if the number of modes in the 
lead or the box is larger than one.

However, as we show now, the temperature $T^*$, below which the two channel
Kondo behavior appears turns out to be extremely small in most 
cases. To see this let us consider  a tunneling junction with a
rough tunneling surface, and a dimensionless 
conductance $g = G/G_0$. Obviously, in this case the  matrix 
elements $t_{mn}$ scale as $t_{mn}\sim \sqrt{g}/N$ so that 
${\rm Tr} \{{\bf t}^\dagger {\bf t}\}\sim g$, and they have random 
sign or phase relative to each-other. The occupation dependent 
back scatterings
 ${\bf q}$ and $\bf \tilde q$ are generated
by Eqs.~(\ref{eq:qscaling}) and (\ref{eq:qtildescaling}), and their 
typical matrix elements can easily be estimated to be of  order
$1/N^{3/2}$. Therefore, the first two terms in
Eq.~(\ref{eq:tscaling}) can be estimated to be smaller than 
$\sim g/N^{2}$, while the matrix elements of the
last term are dominated  by  the term proportional to ${\rm Tr} 
\{{\bf t}^\dagger {\bf t}\}$ and are of  order $\sim g^{3/2} /N$.
Thus for large enough $N\sim M $ one can simply 
substitute ${\bf q} = {\bf \tilde q} = 0$ and the scaling equations 
reduce to:
\begin{equation}
{d{\bf t}\over dx} = - 2 {\bf t}  {\rm Tr}\{ {\bf t t^\dagger}\} 
\end{equation}
Multiplying this equation by ${\bf t^\dagger}$ and taking its trace
we finally arrive at the scaling equation:
\begin{equation}
{ dg\over dx } = - 4\;g^2\; ,
\label{eq:gscaling}
\end{equation}
which  is identical to the one obtained in Ref.~\onlinecite{Schoen}
using the $2N$-channel model. The scaling equation for the effective
field  can also be  expressed in terms of the dimensionless 
conductance:
\begin{equation}
{dh\over dx} = -4g h\;.
\end{equation}
The above two  scaling equations can readily be integrated to obtain:
\begin{eqnarray}
g(x) = {g_0\over 1+ 4g_0 x}\;, 
\label{eq:gsolution} \\
h(x) = {h_0\over 1+ 4g_0 x}\;.
\label{eq:hsolution} 
\end{eqnarray}
Obviously, in this approximation the physics of the SEB is completely 
characterized by the effective field $h$ and the {\em conductance}
of the junction, and the details about the specific structure 
of the junction or the tunneling amplitudes  are unimportant.

It is easy to estimate the crossover scale $T^*$ below which
this approximation breaks down.  The scaling towards 
the two-channel Kondo fixed point is  generated by the second 
order 'coherence' terms in
Eqs.~(\ref{eq:tscaling}-\ref{eq:qtildescaling}), while the third
order terms tend to reduce all couplings to zero.
Therefore the scale $T^*$ is
determined by the condition that the second and third order terms 
in Eqs.~(\ref{eq:tscaling}-\ref{eq:qtildescaling}) 
be of the same order of magnitude, giving
$g\sim 1/N$. Replacing Eq.~(\ref{eq:gsolution}) by its asymptotic 
form $g(x)\approx 1/4x$ we finally obtain
\begin{equation}
T^* \approx E_C \exp\{-C\;N\}\;\phantom{mmm} (N\gg 1)\;,
\label{eq:Tstar}
\end{equation}
where $C$ is a constant of the order of unity.
In view of the experimental values of $E_C$, this scale is 
extremely small for $N>10$, which justifies the use of 
the $2N$-channel model in many experimental setups.

\begin{figure}
\epsfxsize=7.5cm
\epsfbox{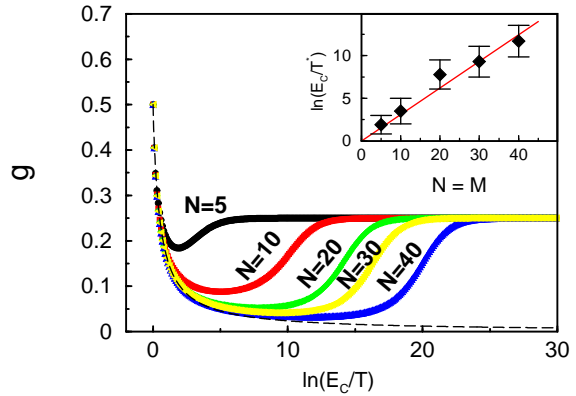}
\vskip0.2truecm
\caption{
\label{fig:gscaling}
Scaling of the conductance of a single electron box at its degeneracy 
point. The conductance  curves have been calculated by 
integrating the full scaling equations
Eqs.~(\protect{\ref{eq:tscaling}-\ref{eq:qtildescaling}}) 
with randomly generated tunneling matrix 
elements. The bare value of the conductance has been fixed to be
$g(0) = g_0 = 0.5$. The conductance follows more and more closely the infinite 
channel formula Eq.~(\protect{\ref{eq:gsolution}}) as the number 
of modes $N=M$  increases. In the inset the logarithm of the crossover 
temperatures corresponding  to the minima   of the conductance curves is 
plotted as a function of the number of modes. It scales linearly with  
the number of modes  in agreement with the arguments presented in the main text.
The large error bars  $\sim \sqrt{N}$ result from the 
fluctuations 
of the tunneling amplitude and indicate that for a small number of 
channels or pinholes substantial sample to  sample deviations 
may occur.}
\end{figure}

\begin{figure}
\epsfxsize=7.5cm
\epsfbox{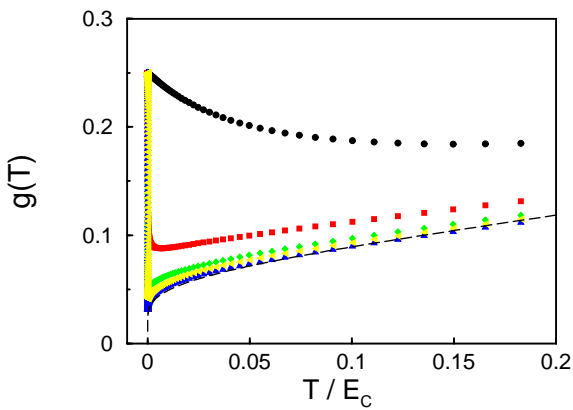}
\vskip0.2truecm
\caption{
\label{fig:linear}
Conductance curves of Fig.~\protect{\ref{fig:gscaling}} on a linear in 
$T$ plot.}
\end{figure}

In Figs.~\ref{fig:gscaling} and \ref{fig:linear}
we show the typical scaling of $g(x)$ for 
various $N$ and $M$ values, obtained  by solving 
Eqs.~(\ref{eq:tscaling}-\ref{eq:qtildescaling}) numerically.
While in the Figures the initial hopping amplitudes have 
been generated completely randomly, similar results have been 
obtained when we used simple model estimates for the $t_{mn}$'s.  
Eq.~(\ref{eq:gsolution}) approximates very nicely 
the conductance above $T^*$ even for rather small channel 
numbers. Below $T^*$, however,  the conductance starts to increase 
until it  reaches its fixed point value $g_{\rm fp} \sim 1$. 
In the inset we show that the scale  $T^*$ decreases exponentially 
with the number of  modes in agreement with Eq.~(\ref{eq:Tstar}).

\section{Discussion of the SET}
\label{sec:set}

To derive the scaling equations for the SET we repeated the 
derivation of the scaling equations of the two-level system 
with a generalized version of the interaction Hamiltonian
(\ref{eq:TLS}). Here we only quote the results. 

The scaling equations become: 
\begin{eqnarray}
{d{\bf t^{\rm (f)}}\over dx}& = & 2({\bf t^{\rm (f)} q^{\rm (f)}} - 
{\bf \tilde q^{\rm (f)} t^{\rm (f)}})
\nonumber \\
& - &
2 {\bf t}^{\rm (f)} \sum_{f^\prime}
\bigl[ {\rm Tr}\{ {\bf t^{\rm (f^\prime)} {t^{\rm (f^\prime)}}^\dagger}\} 
+ 2 {\rm Tr}\{ {\bf q^{\rm (f^\prime)}  q^{\rm (f^\prime)} }\}
\nonumber \\
&+ & 2 {\rm Tr}\{ {\bf \tilde q^{\rm (f^\prime)} \tilde q^{\rm (f^\prime)} 
}\}\bigr]\;,
\label{eq:tscaling2}\\
{d{\bf \tilde q}^{\rm (f)}\over dx} & = &{\bf {t^{\rm (f)}}^\dagger 
t^{\rm (f)}}  - 4 {\bf \tilde q^{\rm (f)}}
\sum_{f^\prime}{\rm Tr}\{ {\bf t^{\rm (f^\prime)} 
{t^{\rm (f^\prime)}}^\dagger}\} \;, 
\label{eq:qscaling2}\\
{d{\bf  q}^{\rm (f)}\over dx} &=& -{\bf  t^{\rm (f)} 
{t^{\rm (f)}}^\dagger}  - 4 {\bf  q}^{\rm (f)}
\sum_{f^\prime}{\rm Tr}\{ {\bf t^{\rm (f^\prime)} 
{t^{\rm (f^\prime)}}^\dagger}\}
\label{eq:qtildescaling2}\;. 
\end{eqnarray}

Similarly to the SEB, the 'incoherent' scaling equations can be 
obtained by dropping all second order 'coherent' terms in the 
equations above. In this way we obtain  for the dimensionless 
conductances $g^{(1)}$  and $g^{(2)}$ of the two junctions 
and the effective field $h$ the following scaling equations: 
\begin{eqnarray} 
{d g^{(1)} \over dx} &=& - 4 g^{(1)} (g^{(1)}+g^{(2)})\;,\\
{d g^{(2)} \over dx} &= &- 4 g^{(2)} (g^{(1)}+g^{(2)})\;,\\
{d h  \over dx}& =& - 4 h (g^{(1)}+g^{(2)})\;.
\end{eqnarray}
These equations can be readily solved to give:
\begin{eqnarray} 
g^{(1)}(x) =  {g^{(1)}_0 \over 1 + 4 (g^{(1)}_0+ g^{(2)}_0) x }\;,
\nonumber\\
g^{(2)}(x) =  {g^{(2)}_0 \over 1 + 4 (g^{(1)}_0+ g^{(2)}_0) x }\;,
\label{eq:incohgs2}\\
h(x) =  {h_0 \over 1 + 4 (g^{(1)}_0+ g^{(2)}_0) x }
\nonumber\;.
\end{eqnarray}
Thus in this approximation the only effect 
of the presence of several leads is to replace  
the dimensionless conductance in the denominator of Eqs.~(\ref{eq:gsolution}) 
and (\ref{eq:hsolution}) by the parallel conductance 
of all junctions  attached to the  island.

\begin{figure}
\epsfxsize=7cm
\epsfbox{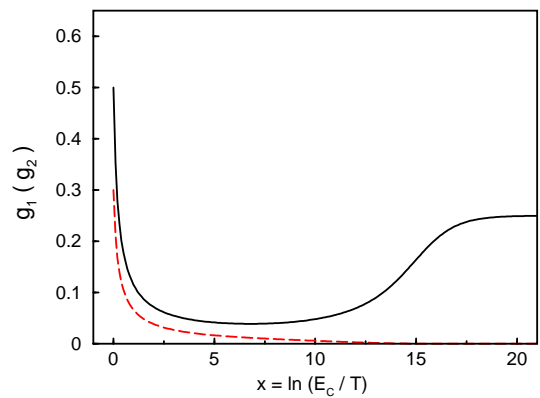}
\vskip0.2truecm
\caption{
\label{fig:setcond}
Conductance of the two junctions of the SET as calculated from 
the solution of the full scaling equations for  $N = M = 15$ channels. 
The tunneling matrix  elements were generated randomly by fixing 
the initial conductances at $g^{(1)} = 0.5$ and $g^{(2)} = 0.3$. 
At low temperatures the system approaches the two-channel Kondo 
fixed point, and one of the conductances scales to a universal value while 
the other scales to zero.}
\end{figure}

In Fig.~\ref{fig:setcond} we show the conductance calculated from the 
solution of the full scaling  equations  
Eq.~(\ref{eq:tscaling2}-\ref{eq:qtildescaling2}) for $N=M = 15$ 
conduction modes. Initially, both conductances  follow 
Eqs.~(\ref{eq:incohgs2}). However, at a temperature 
$T^*\approx 10^{-3} E_C$  a single conduction mode prevails  
 in one of the junctions and starts 
to induce  the two-channel Kondo effect. The conductance of this 
junction approaches a universal value\cite{universal?} characteristic 
to the  two-channel Kondo fixed point while the resistivity of the 
other  junction is suppressed to zero below this temperature. 
Since the tunneling between  the island and this latter lead is an 
irrelevant operator  of dimension $1/2$ and therefore its conductance  
scales as $t^2\sim T$,  the total conductance of the device at 
the degeneracy point scales as  $G \sim T$, in agreement with Matveev's result. 
Note that  the conductance of the other junction is {\em universal}, i.e. 
independent of the number of modes in the junction.

\section{Conclusions}
\label{Conclusions}

In  the present work we studied in detail the physics of 
the single electron box and the single electron transistor 
close to their degeneracy points using renormalization
group methods. In particular, we investigated the effect of 
cross-mode scattering and showed that this is a {\em relevant 
perturbation}, and drives  the system towards a stable 
two-channel Kondo fixed point, a prototype of non-Fermi liquid 
fixed points. At very low temperatures 
we recover Matveev's mapping to the two-channel 
Kondo model:\cite{MatveevJETP,MatveevPRB} In this case at very 
low temperatures  the system  {\em dynamically} selects a single mode 
on the box and  another one  on one of the leads, and all the  other modes 
become irrelevant. This fixed point has an $SU(2)\times SU(2)\times U(1)$
symmetry,\cite{largersymmetry}  where the first symmetry is generated 
dynamically and is connected to the structure of the effective tunneling 
Hamiltonian,   while the second $SU(2)$ symmetry is associated to the real 
spin of the electrons and is responsible for the non-Fermi liquid behavior.
The $U(1)$ symmetry is simply due to charge conservation.

Due to this two-channel Kondo fixed point   
the linear coefficient of the specific heat and the 
capacitance of the SET diverge logarithmically at the degeneracy 
point, $c/T\sim C\sim \ln(T)$,  while the resistivity of the 
device diverges  as $\sim1/T$.\cite{MatveevPRB}

However, as our detailed analysis demonstrates, if the number of 
tunneling modes is larger than one, then a new small energy scale
$T^*\sim E_C \exp(- C N) $ enters into the problem, with $N$ the 
total number of tunneling modes.  Above this scale  coherent processes 
leading to the Kondo physics can be  neglected, and the  
properties of the system are very well 
described  solely  in terms of {\em the conductances} of the 
various junctions. Neglecting  the aforementioned coherent terms we were 
able to re-derive the equations of Ref.~\onlinecite{Schoen}, and 
generalize them for the case of SET.

At this point we have to make an important remark. Our results 
for the two-channel Kondo fixed point 
rely heavily on the fact that the electrons
at temperatures $T$ larger than the level spacing on it
can hardly travel  coherently from one junction to 
another.\cite{AverinNazarov,MatveevPRB} Neglecting this coherent process 
then one of the SET conductances scales to zero and so does the 
total conductance of the SET at the degeneracy point.\cite{MatveevPRB} 

Apart from inelastic scattering, the main  origin of this  loss of coherence is 
in the random scattering  from the wall  of the island and the impurities 
on it.  Any model assuming the existence of  consecutive 
 coherent tunneling events between  different leads   gives an 
essentially different, and very likely in most situations  
unphysical result  at very low temperatures. Indeed, repeating our analysis 
for the model used in Ref.~\onlinecite{KoenigSchoellerSchoen}
we find that at $T=0$ both tunnel junctions of the 
SET have a finite conductance at the degeneracy point, and thus 
the total conductance of the SET also remains finite even at zero 
temperature.  This result is essentially different from 
the one  obtained by Matveev\cite{MatveevJETP,MatveevPRB} 
or the calculations presented here, where coherent 
tunneling processes between different leads are excluded. 

The difference between these two approximations becomes even more 
striking for the case of an $M$-fold degenerate multi-degeneracy 
points of a multi-dot system. In this case, neglecting  the above-mentioned 
coherent  tunneling between different leads, we find that the 
low-temperature physics is again described by the two-channel 
Kondo model. In the opposite case, however, the system would  
scale to another non-Fermi liquid fixed point, 
the $SU(M)\times SU(2)$ Coqblin-Schrieffer fixed point.

How and at what energy scale 
the crossover between these two behaviors happens,  
seems to be presently an open question, which can  only be answered 
by somehow incorporating more  details about  the scattering on the 
impurities and the energy and spatial  dependence of the tunneling 
matrix elements.

Because of the exponential  dependence of $T^*$, even a 
relatively small number of  tunneling modes leads to  an extremely 
small $T^*$ and renders  the non-Fermi liquid behavior in most cases
inaccessible for  the  experimentalists.
To observe the 2-channel  Kondo 
behavior one can use semiconducting  or metallic quantum boxes. 

Semiconducting devices have the advantage that using suitable 
gate electrodes one can realize the idealistic case of a single 
mode contact, and therefore $T^*\sim E_C$ can be achieved. There is a 
serious difficulty, however, since one should keep $E_C$ large 
enough in order to have a measurable Kondo temperature 
while  having a  small level spacing on the island, the latter
 playing  the role of an  infrared cutoff for the Kondo physics. 
The former requirement   immediately sets an upper limit on the size of the 
box $d < e^2/E_C\sim 1000 \AA$, and therefore  
a lower  limit on the mean level spacing $\Delta
\approx 1/(m^* d^2)\sim 0.1-1K$, where we assumed a two dimensional 
electron gas with effective mass $m^*$. This means that
even if one manages to build a semiconducting device
with a Kondo temperature  in the measurable range, 
the level spacing will be  too large to  observe the 
two-channel Kondo behavior in detail. Indeed, even in the very 
recent experiments of  Ref.~\onlinecite{Berman} the ratio 
$E_C/\Delta$ was  in the range $\sim 100$, and consequently 
only some  fingerprints of the two-channel Kondo behavior 
could have been  observed. For smaller islands $E_C$ would 
become larger, however, the ratio $E_C/\Delta$ would get 
even worse. Therefore there seems to be no hope to 
investigate the two-channel Kondo behavior with
semiconducting devices more in detail. 

The other possibility is to prepare {\em metallic} boxes. 
Since these metallic boxes are three dimensional objects
and  $m^*\sim m_e$ (unlike semiconducting devices with $m^* \ll m_e$), 
metallic grains of the size of only $d\approx 100\AA$
may have  quite large $E_C\sim 100-500K$, and very small
mean level spacing on the other hand. Indeed, using STM 
devices to tunnel into metallic drops\cite{STMCoulombBlockade1} 
it was possible to observe 
the Coulomb blockade even at room temperature.\cite{STMCoulombBlockade2}

\begin{figure}
\epsfxsize=7cm
\epsfbox{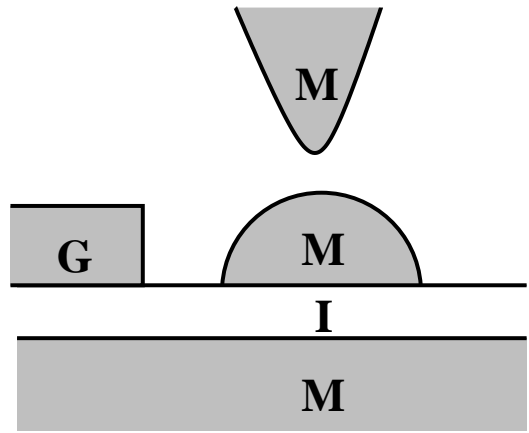}
\vskip0.2truecm
\caption{
\label{fig:stm}
Sketch of an STM setup to observe the two-channel Kondo 
behavior. The external gate electrode over the insulating layer 
is used to tune the metallic droplet to its degeneracy point.}
\end{figure}

The difficulty in this case is connected to the preparation of the junctions. 
As emphasized earlier, one needs practically {\em single mode} or at
 most few mode junctions in order to have $T^*$ large enough, which 
requires atomic size contacts/junctions. 
To establish such a junction we  propose the experimental
setup  illustrated in Fig.~\ref{fig:stm}. 
In the suggested experiment a metallic electrode is covered by a thin
insulating layer, and a metallic drop is deposited on the top of it. 
The atomic size contacts  can be  formed by plugging an STM needle 
into one of the drops  and then gently pulling it out of 
it.\cite{STMsinglemode} 
An additional gate electrode  should be built in the vicinity 
of the drop to control the charging of the island.
The  two-channel Kondo effect would then show up  in the 
gate voltage dependence of the conductance through 
the island.

The biggest difficulty in the experiment above is to establish 
a {\em stable} contact, so that one has enough time to tune 
the drop to its degeneracy point and carry out the measurement. 
It would be much more advantageous to use 
nanotechnology to build atomic size contacts instead of using an
STM, a possibility  which may be not too far away any more.

The authors are grateful to  G. Sch\"on, and M. Devoret for valuable  
discussions. This research has been supported by
the U.S - Hungarian Joint Fund No. 587, grant No. DE-FG03-97ER45640 of the
U.S DOE Office of Science, Division of Materials Research,
and Hungarian Grant Nr. OTKA T026327, OTKA F29236, and
OTKA T029813.


\begin{thebibliography}{99}
\bibitem{footnote} In reality, any roughness of the surface
or non-ideal geometry results in the mixing of these idealistic 
modes. This can be taken into account through a potential scattering 
term in the Hamiltonian, and as far as the mean free path of the 
electrons is larger or comparable to  the width of the junction
does not influence the results obtained.
\bibitem{universal?}
According to the non-perturbative calculations 
of Ref.~\protect{\onlinecite{MatveevPRB}}
the two-channel Kondo fixed point corresponds to perfect transmission 
in a single mode, i.e. $g =1$. Strictly speaking, this value of the 
conductance is out of the reach of the weak coupling theory developped
in this paper, which only predicts that the conductance is of the 
order of one.
\end{thebibliography}
\end{document}